\documentclass[twoside,twocolumn,9pt]{article}
\usepackage{extsizes}
\usepackage[super,sort&compress,comma]{natbib} 
\usepackage[version=3]{mhchem}
\usepackage[left=1.5cm, right=1.5cm, top=1.785cm, bottom=2.0cm]{geometry}
\usepackage{balance}
\usepackage{mathptmx}
\usepackage{amsmath,amssymb}
\usepackage{sectsty}
\usepackage{graphicx} 
\usepackage{lastpage}
\usepackage[format=plain,justification=justified,singlelinecheck=false,font={stretch=1.125,small,sf},labelfont=bf,labelsep=space]{caption}
\usepackage{float}
\usepackage{fancyhdr}
\usepackage{fnpos}
\usepackage[english]{babel}
\addto{\captionsenglish}{%
  
}
\usepackage{array}
\usepackage{droidsans}
\usepackage{charter}
\usepackage[T1]{fontenc}
\usepackage[usenames,dvipsnames]{xcolor}
\usepackage{setspace}
\usepackage[compact]{titlesec}
\usepackage{hyperref}

\usepackage{epstopdf}

\definecolor{cream}{RGB}{222,217,201}

\begin{document}

\pagestyle{fancy}
\thispagestyle{plain}
\fancypagestyle{plain}{
\renewcommand{\headrulewidth}{0pt}
}

\makeFNbottom
\makeatletter
\renewcommand\LARGE{\@setfontsize\LARGE{15pt}{17}}
\renewcommand\Large{\@setfontsize\Large{12pt}{14}}
\renewcommand\large{\@setfontsize\large{10pt}{12}}
\renewcommand\footnotesize{\@setfontsize\footnotesize{7pt}{10}}
\makeatother

\renewcommand{\thefootnote}{\fnsymbol{footnote}}
\renewcommand\footnoterule{\vspace*{1pt}%
\color{cream}\hrule width 3.5in height 0.4pt \color{black}\vspace*{5pt}} 
\setcounter{secnumdepth}{5}

\makeatletter 
\renewcommand\@biblabel[1]{#1}            
\renewcommand\@makefntext[1]%
{\noindent\makebox[0pt][r]{\@thefnmark\,}#1}
\makeatother 
\renewcommand{\figurename}{\small{Fig.}~}
\sectionfont{\sffamily\Large}
\subsectionfont{\normalsize}
\subsubsectionfont{\bf}
\setstretch{1.125} 
\setlength{\skip\footins}{0.8cm}
\setlength{\footnotesep}{0.25cm}
\setlength{\jot}{10pt}
\titlespacing*{\section}{0pt}{4pt}{4pt}
\titlespacing*{\subsection}{0pt}{15pt}{1pt}

\fancyfoot{}
\fancyfoot[CO]{\footnotesize{\sffamily Page \thepage \, of \, \pageref{LastPage}}}
\fancyhead{}
\renewcommand{\headrulewidth}{0pt} 
\renewcommand{\footrulewidth}{0pt}
\setlength{\arrayrulewidth}{1pt}
\setlength{\columnsep}{6.5mm}
\setlength\bibsep{1pt}

\makeatletter 
\newlength{\figrulesep} 
\setlength{\figrulesep}{0.5\textfloatsep} 

\newcommand{\topfigrule}{\vspace*{-1pt}%
\noindent{\color{cream}\rule[-\figrulesep]{\columnwidth}{1.5pt}} }

\newcommand{\botfigrule}{\vspace*{-2pt}%
\noindent{\color{cream}\rule[\figrulesep]{\columnwidth}{1.5pt}} }

\newcommand{\dblfigrule}{\vspace*{-1pt}%
\noindent{\color{cream}\rule[-\figrulesep]{\textwidth}{1.5pt}} }

\makeatother


\newcommand{\vect}[1]{\mathbf{#1}}
\newcommand{\vecIII}[3]{\left(\begin{array}{c}#1\\#2\\#3\end{array}\right)}
\newcommand{\matrixII}[4]{\left(\begin{array}{cc}#1&#2\\#3&#4\end{array}\right)}
\newcommand{\matrixIII}[9]{ \left( \begin{array}{ccc}  #1 & #2 & #3\\  #4 & #5 & #6 \\ #7 & #8 & #9 \end{array} \right)}
\newcommand{\smatrixII}[3]{\left(\begin{array}{cc}#1&#2\\#2&#3\end{array}\right)}

\newcommand{\pd}[2]{\frac{\partial #1}{\partial #2}}
\newcommand{\deriv}[2]{\frac{d #1}{d #2}}
\newcommand{\dif}[0]{\mathrm{d}}
\newcommand{\dS}{\dif S}

\newcommand{\inv}[1]{#1 ^ {-1}}
\newcommand{\invs}[1]{#1 ^ {-2}}
\newcommand{\invf}[1]{\frac{1}{#1}}
\newcommand{\half}[1]{\frac{#1}{2}}
\renewcommand{\exp}[1]{e^{#1}}

\newcommand*\rfrac[2]{{}^{#1}\!/_{#2}}
\newcommand{\brk}[1]{\left( #1 \right)}
\newcommand{\Brk}[1]{\left[ #1 \right]}
\newcommand{\BRK}[1]{\left\{ #1 \right\}}
\newcommand{\splt}[0] {\right. \\ \left.}
\newcommand{\norm}[1]{\left\Vert\vect{#1}\right\Vert}
\newcommand{\abs}[1]{\left|#1\right|}

\newcommand{\ab}{\bar{\vect{a}}}
\newcommand{\bb}{\bar{\vect{b}}}
\newcommand{\aac}{\vect{a}}
\newcommand{\bac}{\vect{b}}
\newcommand{\gb}{\bar{\mathfrak{g}}}
\newcommand{\Chr}[3]{\Gamma^{#1}_{#2#3}}
\newcommand{\Kb}{\bar{K}}
\newcommand{\nrml}{\hat{\vect{n}}}

\newcommand{\figref }[1]{Fig.~\ref{#1}}
\newcommand{\Eqref}[1]{Eq.~\ref{#1}}
\newcommand{\citeSI}[0]{Supplemental Material}
\newcommand{\SFigSwellingTime}[0]{Fig. S1}
\newcommand{\AppendixLambda}[0]{Appendix A}
\newcommand{\AppendixDelta}[0]{Appendix B}

\twocolumn[
  \begin{@twocolumnfalse}

\noindent\LARGE{\textbf{Wrinkling in Sheets with Nonuniform Growth and Bending Rigidity}} \\
\vspace{0.3cm} 

\noindent\large{Ido Levin$^{\ast}$\textit{$^{a}$} and Sarah L. Keller,\textit{$^{b}$}} \\

\noindent\normalsize
    {
    Thin elastic sheets bend easily, leading to mechanical instabilities such as wrinkling. 
    Here we investigate wrinkles at edges of bi-strips, which consist of two thin sheets, one that swells and one that does not, joined side-by-side. It is well known that when bending rigidity is uniform across an isolated bi-strip, swelling results in axisymmetric shapes like a wine bottle: two cylinders of different radii are joined by a smooth transition zone. However, when the bending rigidity of the swollen sheet differs from that of the non-swollen sheet, purely axisymmetric shapes are no longer energetically favorable, and wrinkles arise.
    When the bending rigidity of the non-swollen sheet is essentially infinite, the wrinkles coarsen with distance from the transition zone such that dimensionless wavelengths and widths are related by $\tilde{\lambda} \propto \tilde{w}^{2/3}$. If the bending rigidity of the non-swollen sheet is non-infinite (but~still significantly larger than that of the swollen sheet), then the non-swollen sheet assumes a non-infinite radius of curvature, $R_0$. We find that the wrinkles in this system extend a critical distance, $w_C$ beyond the junction of the two strips and that $w_C \propto R_0$. Local undulations of wrinkles are favorable in this system because they decrease the overall bending energy by allowing the non-swollen sheet to have a larger radius of curvature than would otherwise be dictated by its reference geometry.
    Our results are relevant to a wide range of sheets that experience non-uniform growth, whether in natural systems such as plants or  in synthetic systems such as designed, responsive materials. 
    } \\


 \end{@twocolumnfalse} \vspace{0.6cm}
]

\renewcommand*\rmdefault{bch}\normalfont\upshape
\rmfamily
\section*{}
\vspace{-1cm}


\footnotetext{\textit{$^\ast$}\textit{$^{a}$~Department of Chemistry and Department of Mathematics, University of British Columbia, Vancouver, BC, Canada. Tel: 604.822.3402; E-mail: idolevin@chem.ubc.ca}}
\footnotetext{\textit{$^{b}$~Department of Chemistry, University of Washington, Seattle, WA, USA. }}



\section{Introduction}

Thin sheets are floppy. Because floppy, thin sheets incur a much lower energetic cost to bend than to stretch or compress, they are prone to mechanical instabilities like wrinkling and buckling. Quite literally, wrinkling instabilities are as familiar as the back of your hand. When you pinch the skin on your hand, the wavelength of the wrinkles is determined by an interplay between the bending of the skin (an elastic sheet) and the deformation of the tissue below it (a soft substrate to which the sheet adheres).

\begin{figure}[h!]
    \centering
    \includegraphics[width=8cm]{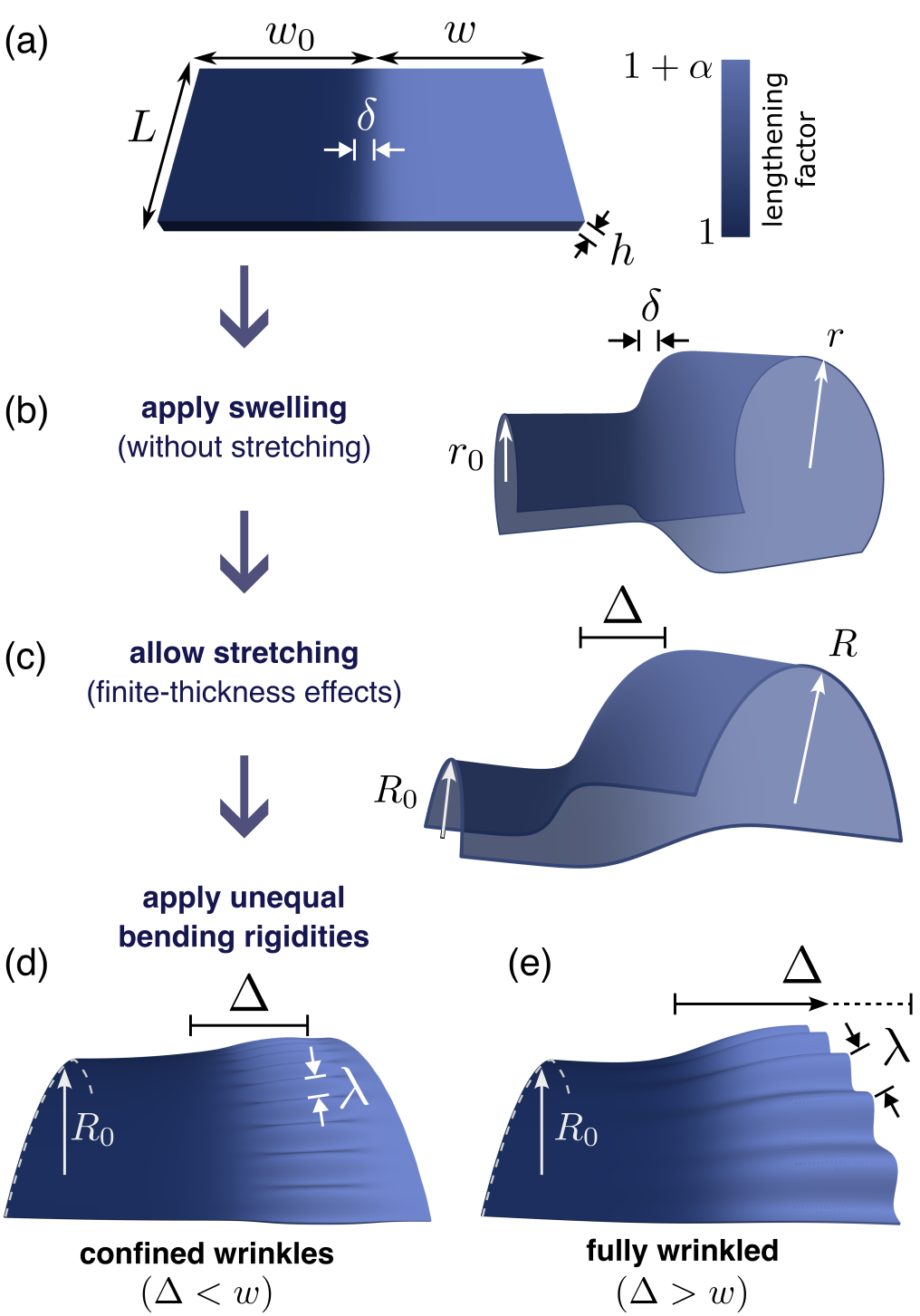}
    \caption{
        Configurations after swelling a bi-strip.
        (a)~The reference geometry is a rectangular sheet with length ($L$), total width ($w_0+w$), and thickness ($h$) subjected to a uniaxial differential swelling field.
        Over most of the sheet, lengths increase by either a factor of 1 (no swelling) or a factor of $1+\alpha$ (maximum swelling). These two regions are joined by a sigmoidal transition over a distance $\delta$.
        (b)~Swelling without stretching results in isometric embedding of the reference geometry: two concentric cylinders are joined by a transition over distance $\delta$.
        (c)~When stretching is allowed, finite thickness effects increase the radii of curvature of both cylinders (now denoted $R_0$ and $R$) and the width of the transition region (now denoted $\Delta$). Stretching energy accumulates in this transition region.
        (d)~The radius of curvature of the non-swollen region can increase further if the two regions have different bending rigidities, which increases the width of the transition region and introduces wrinkles of wavelength $\lambda$.
        (e)~When $\Delta>w$, wrinkles cover the entire swollen region.
    }
    \label{fig:system}
\end{figure}

Here, we create wrinkles by imposing nonuniform growth on thin sheets with nonuniform bending rigidities, also known as "bi-strips". A classic bi-strip is shown in \figref {fig:system}a in which a non-swelling region of width $w_0$, length $L$, and thickness $h$ lies next to a corresponding swelling region of width $w$. The two regions are joined by a sharp transition zone over a distance $\delta$. 
Within this zone, the multiplicative linear swelling factor (or "lengthening") increases sigmoidally from 1 (no swelling) to $1+\alpha$ (maximum swelling).

Bi-strips of this type have been used previously by researchers to explore two separate instabilities: buckling of a wrinkle-free sheet into a large curve of radius $R_0$ \cite{kimThermallyResponsiveRolling2012f,moshePatternSelectionMultiscale2013} or wrinkling of an otherwise flat sheet into small-scale undulations \cite{MoraBoudaoud2006Bi-strip}. 
Here, we investigate the interplay between buckling and wrinkling to understand how large curves and small wrinkles can exist simultaneously in bi-strips. The length scales of both features arise from minimization of the system's stretching and bending energies, without external constraints. Similar energy minimization may underlie the undulations that appear at the edges of many biological structures, such as leaves \cite{sharonGeometricallyDrivenWrinkling2007}, petals \cite{liangGrowthGeometryMechanics2011a,portetRipplesEdgesBlooming2022b} and sea-slugs \cite{yamamotoNaturesFormsAre2021}.

\subsection{Prior work: Large curves in bi-strips with uniform bending rigidity}

Before we describe our results in bi-strips with \emph{nonuniform} bending rigidity, it is helpful to briefly review prior observations of thin bi-strips with \emph{uniform} bending rigidity. These bi-strips \cite{kimThermallyResponsiveRolling2012f} encode incompatible intrinsic geometries with swelling along a single coordinate. The reference geometry inside the transition zone is non-Euclidean (\emph{i.e.}, it has a non-vanishing Gaussian curvature), whereas on either side of the transition zone, it is flat.

If the bi-strip were to remain flat upon swelling, it would experience substantial stretching. 
Instead, the thin bi-strip buckles into an axisymmetric shape resembling a wine bottle: two cylinders of different radii glued together (\figref {fig:system}b) \cite{kimThermallyResponsiveRolling2012f,moshePatternSelectionMultiscale2013}. 
The radius of the swollen region, $r$, is related to the radius of the unswollen region, $r_0$, by $r=(1+\alpha)r_0$. The average radius, $r_\text{avg}$, scales as $\delta$ (the only length scale in the reference geometry).
This scaling fits with the expectation that if swelling varies so slowly across the bi-strip that the transition zone, $\delta$, is infinite, then the sheet remains flat, and $r_\text{avg}$ is infinite, where $r_\text{avg}\equiv\half{1}(r_0+r)$. Even in a curved bi-strip as in \figref {fig:system}b, the two cylinders uncurl far from the transition zone, and the edges lie flat, just as one edge of a long sheet of paper can lie flat when the opposite edge is crimped \cite{baroisHowCurvedElastic2014b}.

When a finite thickness is imposed on the bi-strip, several length scales increase as shown in \figref {fig:system}c: radius $r$ becomes $R$, radius $r_0$ becomes $R_0$, average radius becomes $R_\text{avg}\equiv\half{1}(R_0+R)$, and the length scale, $\delta$, of the transition zone becomes $\Delta$. It follows that the previous relationship $r_\text{avg}\propto\delta$ becomes $R_\text{avg}\propto\Delta$. When thin sheets have finite thickness, a small amount of stretching can occur (see \AppendixLambda). This causes small deviations from the reference geometry and can decrease the bending energy through a variety of mechanisms \cite{efratiBucklingTransitionBoundary2009b,efratiHyperbolicNonEuclideanElastic2011,levinHierarchyGeometricalFrustration2021d}.

\subsection{Prior work: Small wrinkles in bi-strips with nonuniform bending rigidity.~~} 

\begin{figure}[ht]
    \centering
    \includegraphics[width=7.6 cm]{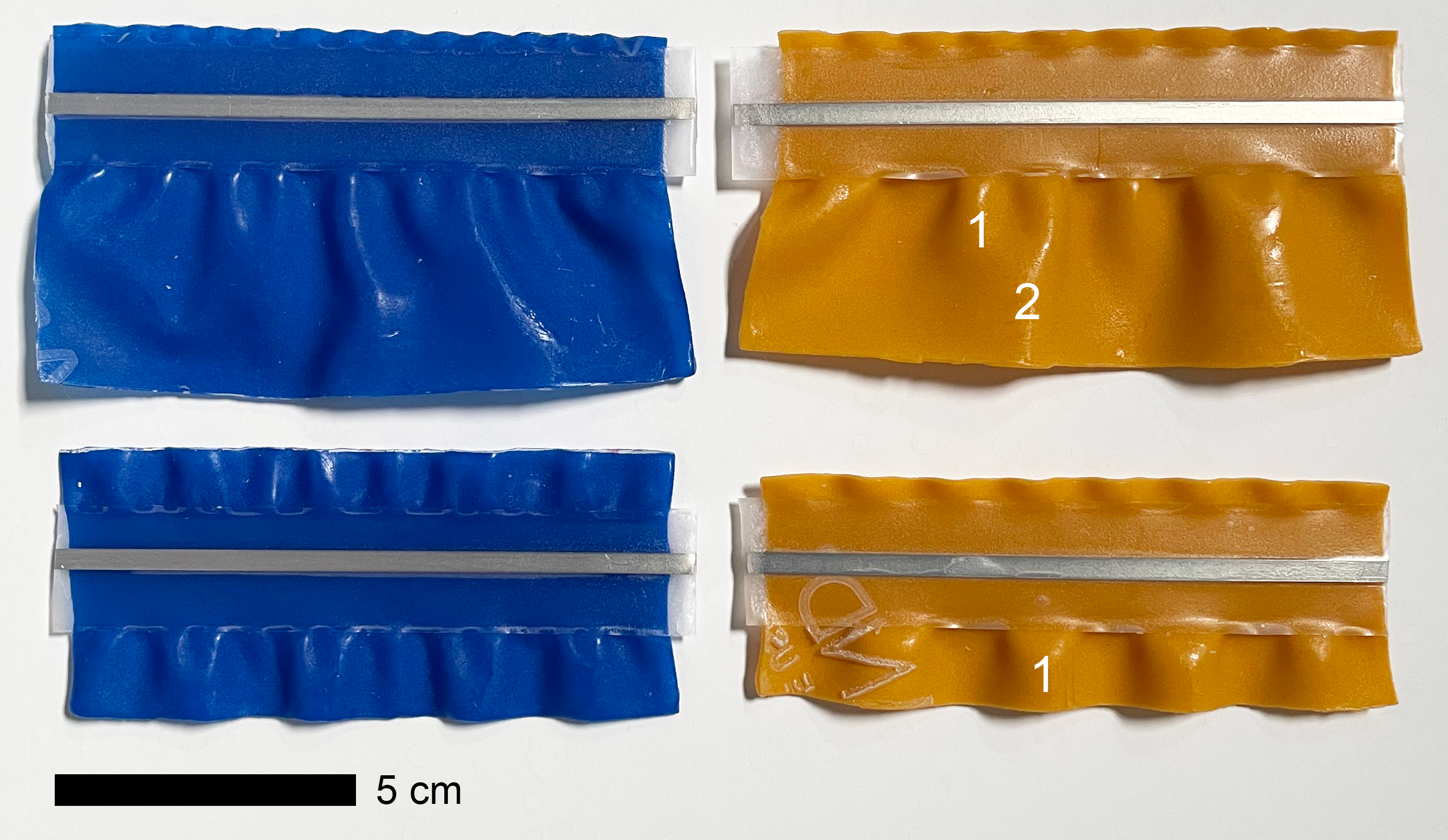}
    \caption{
        Four back-to-back bi-strips. Each has a central, unswollen, flat region of length 9~cm. In swollen regions with a small width, $w$, a single 1st-generation wavelength usually emerges (labeled "1"). In wider swollen regions, wrinkles can merge into a 2nd generation (labeled "2"). 
    }
    \label{fig:4back-to-back}
\end{figure}

Bi-strips with nonuniform bending rigidity have been previously constructed by chemically bonding two strips of gel side by side\cite{MoraBoudaoud2006Bi-strip}. 
When the bending rigidity of the non-swelling strip is essentially infinite, that region remains flat. When the other region swells, the system as a whole cannot relieve elastic energy by buckling into the wine-bottle shape in \figref {fig:system}c. 
Instead, the swelling region acquires wrinkles of wavelength $\lambda$ (\figref {fig:4back-to-back}). Mora and Boudaoud used bi-strips in which the width of the swelling region was up to $\sim$10 times the thickness of the sheet (such that dimensionless width $\tilde{w}\equiv w/h \lesssim 10$)\cite{MoraBoudaoud2006Bi-strip}. Within this restricted regime, only a single wavelength appeared, which was found via linear stability analysis to be $\lambda=3.255 w$\cite{MoraBoudaoud2006Bi-strip}.

\subsection{Prior work: Mechanical instabilities induced by intrinsic geometries}

In the bi-strips above, buckling and wrinkling arose in the \emph{absence} of an external constraint such as compression.
In contrast, there exist many systems in which buckling and wrinkling arise entirely due to external constraints. For example, wrinkles in a patch of pinched skin\cite{danielsonWrinklingHumanSkin1977} (or in polymer layers that mimic skin\cite{SultanBoudaoudBucklingThinGelDisk2008}) are due to an instability in a sheet attached to a soft substrate that is under an externally imposed compression. Similarly, the classic example of Euler buckling in a ruler involves compression of the ruler along its long axis.
Bi-strips belong to a large class of elastic sheets that can be programmed with curved, intrinsic geometries via differential swelling \cite{kleinShapingElasticSheets2007,kimDesigningResponsiveBuckled2012a,aharoniUniversalInverseDesign2018e,siefertBioinspiredPneumaticShapemorphing2019d,hajiesmailiReconfigurableShapemorphingDielectric2019e,nojoomi2DMaterialProgramming2021d,MoraBoudaoud2006Bi-strip,sharonGeometricallyDrivenWrinkling2007,klein2011experimental,white2015programmable,huang2018differential}.
Three-dimensional (3D) shape transformations occur in swollen structures because the swelling field encodes a \emph{reference geometry} that is non-Euclidean and incompatible with a planar sheet.
The final shape minimizes energy through a competition between terms for bending energy and stretching energy.
Thick sheets have high bending rigidity and remain planar after swelling, whereas thin sheets readily bend to minimize stretching.
Whether in programmed elastic sheets without external constraints or in alternative systems with external constraints, transitions from thick, planar sheets to thin, buckled ones follow a similar scaling law \cite{efratiBucklingTransitionBoundary2009b}.

\subsection{New approaches.~~} 

To investigate the interplay between buckling and wrinkling in the absence of external constraints, we create bi-strips with nonuniform bending rigidity that differ in two ways from previous systems. First, we explore wrinkles that persist over distances up to two orders of magnitude larger in $\tilde{w}$ than previously assessed \cite{MoraBoudaoud2006Bi-strip}. We find that small wrinkles that appear near the junction of the two strips coarsen into larger wavelengths further away, with dimensionless wavelengths and widths related by $\tilde{\lambda} \propto \tilde{w}^{2/3}$. 
Second, we explore bi-strips in which the non-swelling region has a radius of curvature $R_0$, corresponding to a non-infinite ratio in bending rigidities (\figref {fig:system}e). We find that buckling and wrinkling appear \emph{concurrently} and that wrinkles do not cover the entire swollen region. Instead, wrinkles extend only a critical distance $w_C \propto R_0$ beyond the junction.

\section{Materials, Methods, Calibration, and Derivations}

\subsection{Materials and Methods}
Latex sheets in eight thicknesses from 0.14 to 0.6~mm were purchased from THERABAND (Akron, OH). Sheets with thicknesses from 0.7 to 1.5~mm, were purchased from McMaster-Carr Supply (Elmhurst, IL). 
To make bi-strips, swelling was blocked in some regions of the sheets by covering both faces with strips of adhesive tape (Scotch magic tape, 3M, St. Paul, MN). The sharp edge of the tape minimizes the width of the transition zone, $\delta$.
A stiff nichrome metal strip (Driver-Harris, Harrison, NJ) of thickness 0.25~mm and width 3.5~mm was affixed to the center of the blocked region (\figref {fig:4back-to-back}). The shape of the metal strip determines the equivalent bending rigidity of the non-swollen region. A flat region corresponds to an infinite bending rigidity, and a region of radius $R_0$ corresponds to high, but non-infinite, bending rigidity.

Exposed regions of bi-strips were swollen by submerging the entire sheet in a bath of liquid paraffin (Blended Waxes, Oshkosh, WI) heated above its melting temperature and then maintained at $60^\circ C$. \cite{tangStretchActivatedReprogrammableShapeMorphing2022,laiTwowayShapeMemory2019}
Wrinkles nucleate at the edge of the adhesive tape and propagate to the free edge of the swelling latex sheet (see Supplemental Video 1).
When the sheet is warm and the latex is fluid, the wrinkles have a soft edge mode: they can be translated by running a finger along the edge of the sheet.
When the sheet is removed from the bath, it cools, and the paraffin in the latex solidifies, ``locking'' the sheet in its current shape (\figref {fig:4back-to-back}).   

\subsection{Calibration}

\begin{figure}[h!]
    \centering
    \includegraphics[width=8.6cm]{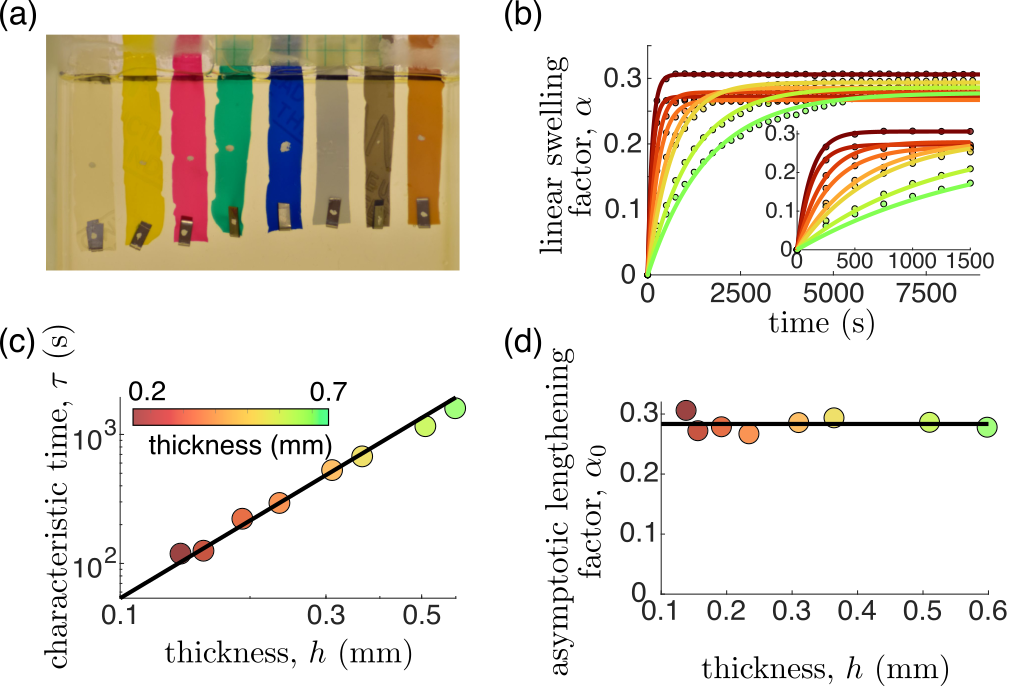}
    \caption{
        Measurement of the lengthening factor, $\alpha(t)$.
        (a)~Photograph of latex sheets of different thicknesses submerged in molten paraffin. Each sheet began as an 8 × 35~mm rectangle.
        (b)~Lengthening as a function of time for the sheets in panel a.
        Each swelling curve was fit to an exponential function with an asymptotic lengthening factor, $\alpha_0$, and a characteristic time, $\tau$.
        (c)~Characteristic times from the exponential fits as a function of the thickness, $h$, of each latex sheet.
        The timescales follow $\tau\propto h^2$, as expected from the diffusion of a solvent into a porous material.
        (d)~Asymptotic lengthening factors are independent of the thicknesses of the latex sheets. The mean lengthening factor is $\alpha_0 \approx 0.28$, with a standard deviation of $0.02$.
    }
    \label{fig:calibration}
\end{figure}

Exposed regions of sheets swell uniformly, with an asymptotic linear swelling factor, $\alpha$, of about 30\% (\figref {fig:calibration}). 
To measure $\alpha$, rectangular strips (approximately $8\times35$ mm) were cut from latex sheets of different lengths, and fiduciary marks were made on the sheets with a white correction pen.
Small metal masses (0.1~g) were attached to the bottom edge of the strips to keep them straight. 
No measurable stretching resulted from the masses.
The strips were attached to an upper railing, then were submerged in molten paraffin (\figref {fig:calibration}a).
A D5500 camera (Nikon, Tokyo, Japan) with a 100~mm macro lens (Tokina, Tokyo, Japan) at the front of the bath recorded an image every $10$ s over 2.5 hours.

Lengthening factors $\alpha(t)$ were found by mapping trajectories of two fiducial marks in each strip over the time lapse, using the Tracker analysis tool \cite{brownTrackerVideoAnalysis2022}.
Measured lengthening factors in \figref {fig:calibration}b were fit to exponential curves of the form:
\[
\alpha(t) = \alpha_0 \brk{1-\exp{-\frac{t}{\tau}}}.
\]
Fits yielded characteristic times of swelling, $\tau$, and the asymptotic lengthening factor, $\alpha_0$.
Characteristic times scaled as $\tau(h) \propto h^2$ (where $h$ is thickness), as expected for solvent diffusion into a thin porous material (\figref {fig:calibration}c).
The mean asymptotic lengthening factor was $\sim 0.28$, with standard deviation $0.02$ (\figref {fig:calibration}d).
Although the sheet thickness affects the time for the latex sheet to swell completely, it does not affect the asymptotic lengthening factor.

\section{Results and Discussion}

\subsection{Wrinkles at edges of flat sheets.~~} 

In this section, we consider bi-strips in which the non-swelling region is flat as in \figref {fig:4back-to-back}. As swelling progresses, the bi-strip cannot relieve elastic energy by bending into the wine-bottle shape in \figref {fig:system}c. Instead, wrinkles  
appear. During the swelling process, 1st-generation wrinkles with a single, uniform wavelength can coarsen into 2nd- and 3rd-generation wrinkles (\figref {fig:4back-to-back}).

\begin{figure}[h!]
    \centering
    \includegraphics[width=8.6cm]{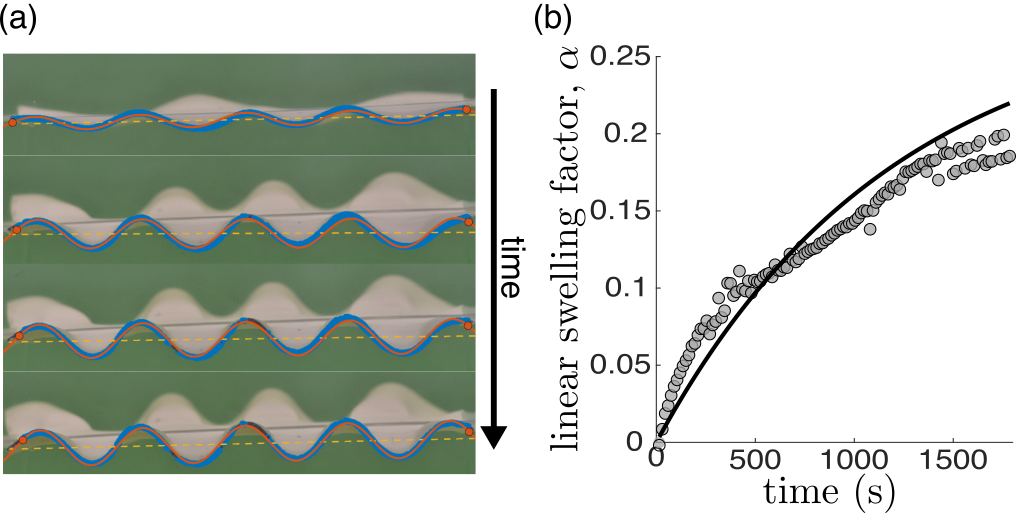}
    \caption{
        Profiles of 1st-generation wrinkles that appear in a bi-strip as it swells in liquid paraffin.
        (a)~Photographs from the side, through time, of one edge of a back-to-back bi-strip. The sheet is flat at $t = $~0~s (not shown). Over time, wrinkles grow. The bottom image corresponds to $t = $~80~s. The profile of the edge (blue) is fit with a sine wave (red) to extract the linear swelling factor, $\alpha(t)$.
        (b)~The linear swelling factor extracted from Panel a (gray circles) is fit well by an exact prediction from the calibrated swelling factor in \figref {fig:calibration}, which means there is no significant stretching in 1st-generation wrinkles.
    }
    \label{fig:profile}
\end{figure}

Our goal is to relate the wavelength of the wrinkles, $\lambda$, to the width of the swollen region, $w$. For higher accuracy, we restrict our measurements to long bi-strips, with $L>1.5\lambda$. In \figref{fig:profile}, we confirm a key assumption: nearly no stretching occurs along the path length of a uniform wrinkle. As the flat bi-strip in \figref{fig:profile} swells over time, the profile of its edge fits a sine wave with a path length prescribed by the lengthening factor in \figref {fig:calibration}.

\begin{figure}[ht]
    \centering
    \includegraphics[width=8.6cm]{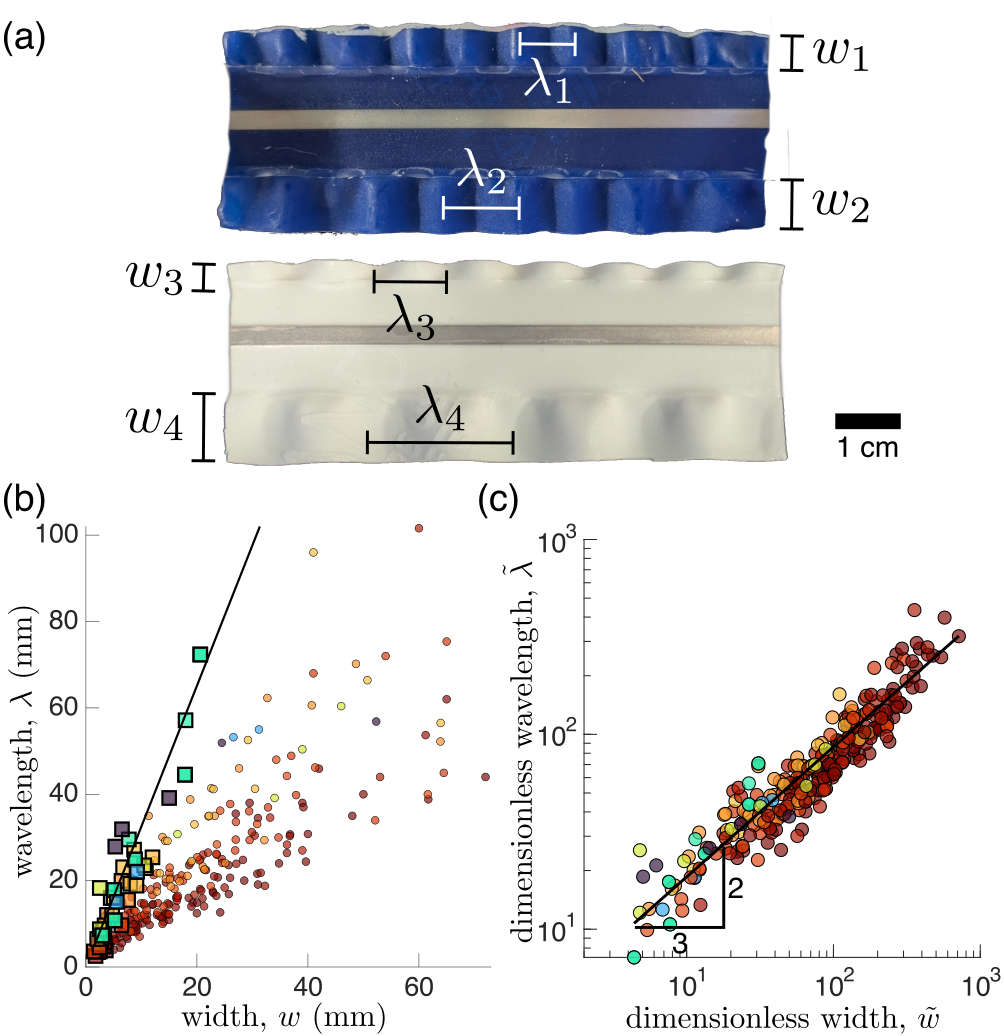}
    \caption{
        Scaling of wrinkles in back-to-back bi-strips in which the unswollen central region is flat. 
        (a)~Wrinkles in latex sheets with thickness 0.31 mm and length 9~cm (blue sheet, top) and 0.51 mm (white sheet, bottom). 
        The sheets have non-swollen central regions and swollen edge regions of widths $w_1$, $w_2$, $w_3$, and $w_4$.
        (b)~Measured wavelength versus width of swollen regions. Color corresponds to the thickness of the sheet before swelling. Squares denote 1st-generation wrinkles fit by $\lambda = 3.255w$ (black line)\cite{MoraBoudaoud2006Bi-strip}. Circles denote wrinkles that have undergone further coarsening. 
        (c)~Data from coarsened 2nd- and 3rd-generation wrinkles collapse to the predicted power law of $\tilde{\lambda} \propto \tilde{w}^{2/3}$ when replotted as dimensionless wavelength $\tilde{\lambda}\equiv\lambda/h$ versus dimensionless width $\tilde{w}\equiv\alpha^{-1/4}w/h$.
    }
    \label{fig:flat}
\end{figure}

It is clear by eye from \figref {fig:flat}a that the wavelength of 1st-generation wrinkles increases with both the initial width of the swelling region and the thickness of the sheet. As shown by other authors\cite{MoraBoudaoud2006Bi-strip}, 1st-generation wrinkles follow a scaling of $\lambda=3.255w$ (\figref {fig:flat}b). Our data extend the range over which that function fits. However, subsequent generations of coarsened wrinkles are poorly fit by $\lambda \propto w$ (\figref {fig:flat}b). Coarsening was not observed in previous studies of flat bi-strips because the widths were smaller\cite{MoraBoudaoud2006Bi-strip}

When data for 2nd- and 3rd-generation wrinkles are replotted in terms of dimensionless wavelength ($\tilde{\lambda}\equiv \lambda/h$) and dimensionless width ($\tilde{w} \equiv \alpha^{-1/4} w/h$) as in \figref {fig:flat}c, the data collapse onto a power law of:
\begin{equation}
    \tilde{\lambda} \propto \tilde{w}^{2/3}
    \label{eq:lambda_flat}
\end{equation}

In bi-strips, coarsening that occurs through a doubling of the wavelength reduces the energy of a uniformly wrinkled configuration (see \AppendixLambda). As in hanging curtains, the following equation applies\cite{vandeparreWrinklingHierarchyConstrained2011}:
\begin{equation}
    \lambda(x) \sim \alpha^{-1/6} h^{1/3} x^{2/3}
    \label{eq:curtain}
\end{equation}

The power law in \Eqref{eq:lambda_flat} agrees with \Eqref{eq:curtain}, when $x$, the distance from the compressed edge, is replaced by $\tilde{w}$
\cite{vandeparreWrinklingHierarchyConstrained2011}. 

An analogy between bi-strips and hanging curtains is not obvious. A key difference between the two systems is that wrinkles in bi-strips arise from their non-Euclidean intrinsic geometry, whereas hanging curtains are subject to external constraints. Another important difference is that the initial wavelength of wrinkles in the bi-strip in \figref {fig:flat} arises because the higher bending rigidity of the unswollen region makes axisymmetric configurations as in \figref{fig:system}c energetically unfavorable. In curtains, the wavelength is set by an external constraint. 
Many other systems involve externally constrained thin sheets, including thin sheets stretched along one axis \cite{cerdaGeometryPhysicsWrinkling2003}, curved shells on flat fluid surfaces \cite{kingElasticSheetLiquid2012b}, Mylar balloons \cite{siefertGeometryMechanicsInextensible2020b}, and curved sheets compressed between plates \cite{hureStampingWrinklingElastic2012,stein-montalvoBucklingGeometricallyConfined2019f,duffyLiftingLoadingBuckling2023}.
In these cases, the constraints cause accumulation of elastic energy in the form of bending and stretching (a term that includes compression), and this energy can be lowered when the sheet wrinkles.
When a sheet is deformed in only one direction, parallel wrinkles can emerge, and their wavelength can be calculated as the only unknown arising from a competition between bending and stretching terms. Low bending rigidities in thin sheets result in highly curved wrinkles.
More complex wrinkle patterns \cite{breidCurvaturecontrolledWrinkleMorphologies2013,aharoniSmecticOrderWrinkles2017,wangCurvatureRegulatedMultiphasePatterns2023b} require more complex calculations to solve for wavelengths and orientations \cite{aharoniSmecticOrderWrinkles2017,tobascoExactSolutionsWrinkle2022a}. 

Whether a system is subject to a non-Euclidean intrinsic geometry or external constraints, once an initial, uniform wavelength $\lambda$ is set, the total accumulated elastic energy can be reduced by allowing coarsening. 
In both cases, coarsening occurs via a doubling of the wavelength as in \figref {fig:4back-to-back}.
Each doubling event results in local stretching confined to small areas of the sheet in which two wrinkles merge into one. By considering these areas, coarsening dynamics can be found by comparing the accumulated stretching energy to the reduced bending energy (see \AppendixLambda).
The total energy due to local stretching is small.
Each doubling event can occur only if it reduces bending energy more than it contributes stretching energy.
Because local bending incurs little energy, the amount of stretching accumulated in each doubling event is limited.

\subsection{Wrinkles in curved sheets.~~} 

Next, we consider bi-strips in which the bending rigidity of the non-swollen region is finite, but is still larger than that of the swollen region. Intuition about how these sheets behave can be gained by comparing the limiting cases of \figref {fig:system}c (in which bi-strips with uniform bending rigidity form wrinkle-free, joined cylinders) and \figref {fig:flat}a (in which bi-strips with flat regions form wrinkles at their edges). In an intermediate case, the non-swollen region would bend with a new radius $R_0$ that lies between the former $R_0$ in \figref {fig:system}c and the infinite radius in \figref {fig:flat}. When the swollen region is attached to an unswollen region with the new, intermediate radius $R_0$, it may wrinkle to alleviate its elastic energy as shown in \figref {fig:system}d. 

\begin{figure}[ht]
    \centering
    \includegraphics[width=8.6cm]{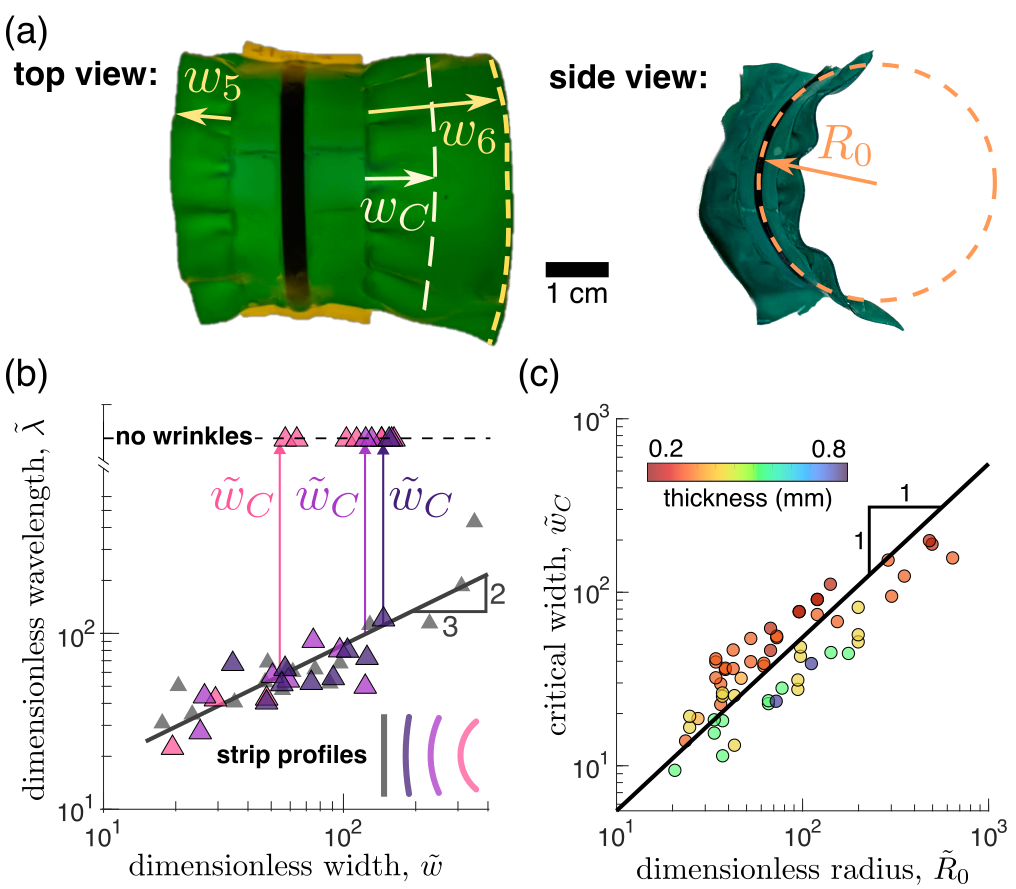}
    \caption{
        Scaling of wrinkles in back-to-back bi-strips in which the unswollen region is curved to a prescribed radius of $R_0$. 
        (a)~Wrinkles in the swollen region are either absent (as in \figref {fig:system}c) or persist to a critical width, $w_C$. The top view shows critical widths that are both longer than the swollen region ($w_C \geq w_5$) and shorter than it ($w_C<w_6$). 
        (b)~Wrinkles at the free edge of curved strips of thickness 0.23 mm follow a power law scaling of $\tilde{\lambda} \propto \tilde{w}^{2/3}$ as in \figref {fig:flat}c when $\tilde{w_C}>\tilde{w}$. The wrinkles disappear when $\tilde{w_C}<\tilde{w}$ (upward arrows). Data for prescribed radii of 7.5~mm (light pink), 17~mm (lavender), and 37.5~mm (violet) are plotted as large triangles. Small gray triangles denote data for bi-strips of the same thickness in which the unswollen region is held flat. 
        (c)~Measured critical widths and prescribed radii collapse when plotted as dimensionless values of $\tilde{w}_C \propto \tilde{R}_0$.
    }
    \label{fig:curved}
\end{figure}

Physically, this intermediate system can be achieved without external constraints by increasing the bending rigidity of the non-swelling region of the bi-strip, for example by making it thicker or increasing its width. An equivalent method is to affix a metal bar to the non-swelling region of the bi-strip, bend the bar into an arc of intermediate radius $R_0$, and then allow the other regions to swell (\figref {fig:curved}a). Because the bending rigidity of the metal bar is much greater than that of latex, its radius does not change when the bi-strip swells.

We find that wrinkles do indeed form in these intermediate, curved bi-strips, and that coarsening of the wrinkle wavelengths follows the same scaling of $\tilde{\lambda} \propto \tilde{w}^{2/3}$ as in flat bi-strips. However, the wrinkles do not cover the entire swollen region (\figref {fig:curved}a). Instead, they stop abruptly at a critical width from the edge of the unswollen region, $w_C$.
We can estimate the critical width by reasoning that wrinkles appear in areas of the bi-strip that cannot relieve stretching energy. Because stretching energy due to swelling accumulates in the transition zone, we expect the critical width to approximately span the transition zone: $w_C\approx \Delta$.
To estimate $\Delta$, we reasoned that $R_\text{avg} \propto \Delta$ (see \AppendixDelta). In turn, from \figref {fig:curved}a, we observe that $R_\text{avg} \approx R_0$. Taken together, these yield:

\begin{equation}
    w_C \propto R_0
    \label{eq:wC_and_R0}
\end{equation}

Dividing by the thickness of the sheet yields the dimensionless value $\tilde{R}_0\equiv R_0/h$, where
\begin{equation}
    \tilde{w}_C \propto \tilde{R}_0\
    \label{eq:dimensionless_wC_and_R0}
\end{equation}
Our experimental data do indeed collapse to this power law, as shown in \figref {fig:curved}c.

More generally, \figref {fig:curved} shows that when sheets change shape due to an incompatibility of their intrinsic geometry, they can exhibit both buckling \emph{and} wrinkling.
Although a global buckled solution may be energetically favorable in some cases, in other cases nonuniform bending rigidity may result in localized wrinkles.
Depending on the curvature of the buckled solution, the wrinkles may be confined to specific regions.

\section*{Conclusions}
In this work, we characterized growth-induced wrinkles at the edges of thin bi-strips, a well-controlled experimental system without external constraints. 
The bi-strips represent an actuable material with emergent, small-scale features.
Similar edge wrinkles are common in natural systems (as in leaves and petals) and lend themselves to quantitative analysis via geometric analysis of the configuration. The wrinkles arise due to a competition between bending and stretching, a known source of mechanical properties in buckled sheets \cite{schenkZeroStiffness2014,levinAnomalouslySoftNonEuclidean2016p,arieliMechanicalDesignPrinciples2024}.
In bi-strips, the wrinkling instability can reduce the total energy of bending and stretching by allowing the system to approximate its reference geometry.
A decrease in overall bending energy might seem counterintuitive, given that wrinkles \emph{increase} the \emph{local} bending energy, but by confining wrinkles to specific regions, the \emph{total} bending energy of the sheet is reduced.
Furthermore, the coexistence of wrinkles and buckling may lead to spatial confinement of the wrinkles.
This confinement arises from their geometrical origin: wrinkles emerge locally to mitigate a mismatch between the reference metric and an unwrinkled configuration that would otherwise lead to stretching.

Our results represent a step toward modeling complex patterns of sharp growth gradients, which can induce intricate large-scale morphologies \cite{therien-aubinShapeTransformationsSoft2015b}.
When materials swell, their mechanical properties change.
Therefore, differential swelling generally results in nonuniform bending rigidity.
This effect is typically either neglected in the literature or regarded as an unwanted artifact.
Our results suggest a way to enrich the types of shapes available to shape-morphing systems by exploiting differences in bending rigidities.
Complementary systems that hold promise in this area are hydrogels that can be rigidified \cite{therien2013multiple} and elastomer sheets that are swollen by inter-diffusing polymers \cite{pezzulla2015morphing}.


\section*{Author contributions}

I.L. developed the concept, method, and theory; collected and analyzed data; and produced data tables and figures. I.L. and S.L.K. wrote and edited the manuscript.

\section*{Conflicts of interest}
There are no conflicts to declare.

\section*{Data availability}

The data supporting this article have been included as part of the Supplementary Information.

\textbf{Description of Supplementary Movies:} Movie 1 – Videos of three back-to-back latex bi-strips of different thicknesses (0.31, 0.51 and 0.60 mm) swelling in liquid paraffin. The duration of the video is 30 minutes and the playback speed is $\times150$.

\textbf{Description of Supplementary Data:} All data used in this research are included in data\_file.xlxs. Data are labeled and sheet names corresponds to each experiment. 

\section*{Acknowledgements}
I.L. was supported by the Washington Research Foundation. S.L.K. acknowledges funding from the National Science Foundation MCB-2325819.

\section*{Appendix A: Derivation of scaling of wrinkle wavelengths}

The wavelength of wrinkles is determined by a local competition between bending and stretching.
Derivation of a scaling relationship for coarsening of this wavelength follows a line of reasoning similar to that of curtains gathered and hung on a flat railing \cite{vandeparreWrinklingHierarchyConstrained2011}.

We define the wavelength and amplitude of wrinkles to be $\lambda(x)$ and $A(x)$, respectively.
In \figref{fig:system}, the $x-$direction runs from left to right in the page, and the $y-$direction is into the page. 
Wrinkles arise due to an increased reference length $\bar{L}_\alpha(x) = \alpha(x) L$, where $\alpha(x)$ is the \emph{local} swelling factor, and $L$ is the original length of the sheet.
Length $L_\alpha(x)$ is a path length that follows the shape, z(y), of the wrinkles:
\[
L_\alpha = \int_{0}^{L} \sqrt{1+\brk{\pd{z}{y}}^2} \,dy
\]
If we temporarily constrain our discussion to only the wrinkles within a single generation (and do not include the doubling transition between wrinkles), then \figref {fig:profile} shows that we can assume no stretching in the wrinkles, such that $\bar{L}_\alpha(x) = {L}_\alpha(x)$. When there is no measurable stretching in wrinkles, the swelling factor $\alpha(x)$ becomes:
\[
\alpha(x) = \frac{\bar{L}_\alpha}{L} =\frac{L_\alpha}{L} = \frac{1}{L}\int_{0}^{L} \sqrt{1+\brk{\pd{z}{y}}^2} \,dy \sim \brk{\frac{A(x)}{\lambda(x)}}^2
\]
Rearranging this equation yields:
\[
A(x) \sim \lambda(x) \sqrt{\alpha(x)}
\]
Therefore, in the absence of stretching, the amplitude of wrinkles is constrained by the swelling factor and the wavelength. This equation is analogous to that for the amplitude of a buckled beam, $A = \lambda \sqrt{strain}$, when $strain$ is $\alpha$.
We assume that the amplitude of the wrinkles changes slowly, such that the dominant curvature is due to the wrinkles themselves:
\[
\kappa(x) \sim \frac{A(x)}{\lambda^2} \sim \frac{\sqrt{\alpha(x)}}{\lambda(x)}
\]

Now we consider the doubling transition between wrinkles. Although the energy required to bend a thin sheet is small, it is nonzero for a sheet of finite thickness ($h \neq 0$). The bending energy density, $\mathcal{E}_B$, generally scales as $\mathcal{E}_B\sim h^3 \kappa^2$, which means that the bending energy density of the wrinkles scales as $\mathcal{E}_B \sim h^3 \alpha \lambda^{-2}$. The wavelength of the wrinkles coarsens over a scale $D$ that varies with each generation. The bending energy, $E_B$, associated with each generation of wrinkles is:
\[
E_B \sim D \mathcal{E}_B \sim D h^3 \alpha \lambda^{-2}
\]

Coarsening of wrinkles reduces a small amount of bending energy, at the cost of a small amount of stretching energy.
As noted above, the amplitude of the wrinkles changes slowly. This change in amplitude results in a small mismatch between $L_\alpha$ and $\bar{L}_\alpha$.
The associated strain, $\epsilon$, scales as $\epsilon \sim (A/D)^2 \sim \alpha \lambda ^2 D^{-2}$.
Therefore, the stretching energy, $E_S$, associated with each generation of wrinkles is:
\[
E_S \sim h D \epsilon^2 \sim h \alpha^2 D^{-3} \lambda ^4
\]

Minimizing the total energy, $E=E_B+E_S$, with respect to $D$, yields $D\sim \alpha^{1/4} h^{-1/2} \lambda ^{3/2}$.
We rearrange this relationship to yield $\lambda \sim \alpha^{-1/6} h^{1/3} D^{2/3}$. Assuming that $\lambda(x)$ changes smoothly (in order to obtain a scaling law rather than a discrete doubling), we substitute $D$ for the continuous variable $x$ to find:
\begin{equation}
    \lambda(x) \sim \alpha^{-1/6} h^{1/3} x^{2/3}
\end{equation}

\section*{Appendix B: Derivation of scaling of the transition region}

\begin{figure}[h!]
    \centering
    \includegraphics[width=6.5cm]{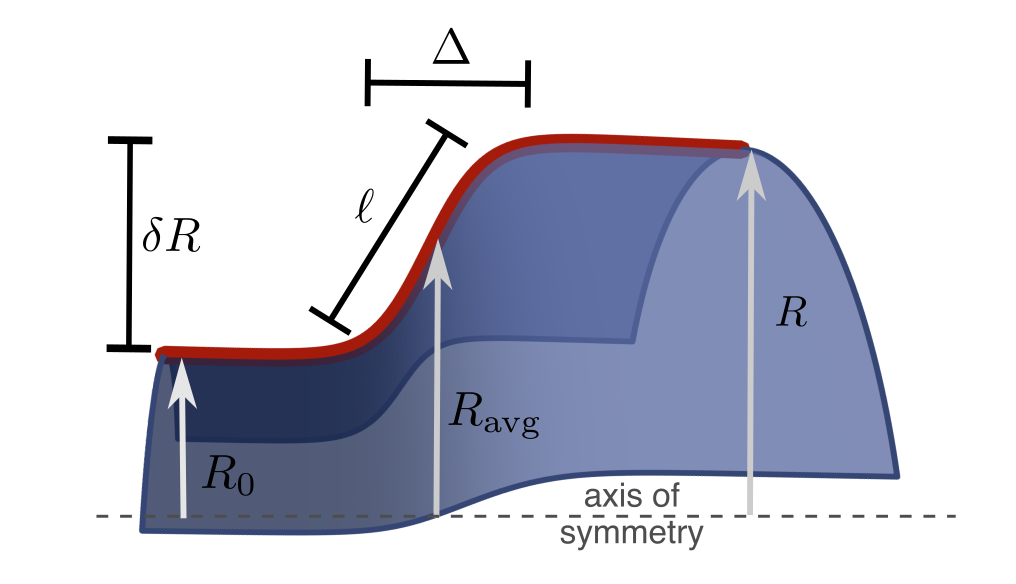}
    \caption{
        An axisymmetric bi-strip.
        A cylinder of radius $R_0$ and a cylinder of radius $R=(1+\alpha) R_0$ are joined over a transition region of width $\Delta$.
        The figure also shows the average radius, $R_\text{avg} = 1/2(R_0+R)$, the difference between the radii, $\delta R = R-R_0$, and the length of the transition region in the material coordinate, $\ell$.
    }
    \label{figs:geometry}
\end{figure}

Here we analyze the shape in \figref {figs:geometry} to show that the width of the transition region, $\Delta$, is proportional to the average radius. An analogous argument appears in \cite{kimThermallyResponsiveRolling2012f}. In \figref {figs:geometry}: 

\begin{itemize}
    \item $\Delta$ is the length of the transition zone in the direction of the axis of symmetry, and $\ell$ is the length of the transition zone in material coordinates.
    \item $\alpha$ is the lengthening factor, where $1+\alpha=R/R_0$.
    \item $R_\text{avg} = 1/2\,(R_0+R)$ is the average radius.
    \item $\delta R = R-R_0$ is the difference between the radii.
\end{itemize}

The length $\ell$, obeys $\ell\sim(1+\alpha) \Delta$.
Furthermore, $\ell$ is approximately the hypotenuse of a triangle where $\ell^2\sim \Delta^2 + \delta R^2$. Subsituting for $\ell$ yields $[(1+\alpha)^2-1]\Delta^2 \sim \delta R^2$, where $(1+\alpha)$ has the same value in all experimental samples and can be treated as simply a number such that 
\[
\Delta \sim \delta R.
\]

Next, definitions for $\alpha$ and $R_{avg}$ give $1+\alpha=R/R_0=\frac{R_\text{avg}+\delta R/2}{R_\text{avg}-\delta R/2}$. Again, $(1+\alpha)$ is simply a number. Rearranging the equation gives $R_\text{avg} \propto \delta R$.
Substituting the previous above that $\Delta \sim \delta R$ yields $\Delta \sim R_\text{avg}$.



\balance


\bibliography{references} 
\bibliographystyle{rsc} 

\end{document}